\documentclass[pre,aps,twocolumn]{revtex4}

\usepackage{times}
\usepackage{amsmath}
\usepackage{graphicx}
\usepackage{amssymb}
\usepackage{color}

\newcommand{\wb}{\omega}

\begin{document}

\title{An energy criterion for the spectral stability of discrete breathers}

\author{Panayotis\ G.\ Kevrekidis}
\affiliation{Department of Mathematics and Statistics, University of
Massachusetts, Amherst, MA 01003-9305, USA}

\author{Jes\'us Cuevas--Maraver}
\affiliation{Grupo de F\'{\i}sica No Lineal, Departamento de F\'{\i}sica Aplicada I, Universidad de Sevilla. Escuela Polit{\'e}cnica Superior, C/ Virgen de \'Africa, 7, 41011-Sevilla, Spain \\ Instituto de Matem\'aticas de la Universidad de Sevilla (IMUS). Edificio Celestino Mutis. Avda. Reina Mercedes s/n, 41012-Sevilla, Spain}

\author{Dmitry E. Pelinovsky}
\affiliation{Department of Mathematics, McMaster University, Hamilton, Ontario, Canada, L8S 4K1 \\
Department of Applied Mathematics, Nizhny Novgorod State Technical University, 24 Minin Street, Nizhny Novgorod, Russia}

\begin{abstract}
Discrete breathers are ubiquitous structures in nonlinear anharmonic models
ranging from the prototypical example of the Fermi-Pasta-Ulam model
to Klein-Gordon nonlinear lattices, among many others. We propose
a {\it general} criterion for the emergence of
instabilities of discrete breathers analogous to the
well-established Vakhitov-Kolokolov criterion for
solitary waves. The criterion involves the change
of monotonicity of the discrete breather's energy as a function of
the breather frequency. Our analysis suggests and numerical results
corroborate that breathers with increasing (decreasing) energy-frequency dependence are generically
unstable in soft (hard) nonlinear potentials.
\end{abstract}
\maketitle

{\it Introduction.}
Discrete breathers, also referred to as intrinsic localized modes,
are time-periodic and exponentially localized
in space coherent structures that have been extensively studied
over the last three decades; see, e.g.,~\cite{aubry,FlachPR2008}.
Their relevance has been recognized not only
theoretically but, importantly, via physical experiments in
areas as diverse as Josephson junction
arrays~\cite{TriasPRL2000,BinderPRL2000},
micro-mechanical cantilever arrays \cite{SatoPRL2003,SatoC2003},
coupled antiferromagnetic
layers \cite{SchwarzPRL1999},  electrical
transmission lines \cite{EnglishPRE2008}, halide-bridged transition metal
complexes~\cite{swanson},
and  torsionally-coupled pendula~\cite{cuevas} among numerous
others.
Remarkably, their areas of purview continue to grow with a
recent example being, e.g., granular crystals in material
science~\cite{BoechlerPRL2010,jinkyu}. Essentially, it is
recognized that broad classes of nonlinear dynamical lattices,
including the paradigmatic (for nonlinear science) case of
the Fermi-Pasta-Ulam (FPU) problem~\cite{fpuref,Chaos},
as well as that of Klein-Gordon (KG)
chains support a plethora of such states.

Since the energy function is typically the only conserved quantity for
the FPU and KG chains, stability criteria that are well-established
for solitary waves, such as the famous Vakhitov--Kolokolov (VK) slope
condition~\cite{VK75},
{\it do not apply} to classify their stability. As a result,
most studies of stability of discrete breathers chiefly rely
on numerical experiments
and a qualitative analysis of eigenvalues in the Floquet-Bloch spectra
of the time-periodic linearization operators~\cite{MS98,MAF98,cls,Noble}.
Some analytical results on the stability of discrete breathers for KG lattices were obtained
by using the limit of small coupling between nearest lattice sites, typically
referred to as the anti-continuum (AC) limit~\cite{MA94}. In this limit,
asymptotic stability of the fundamental (single-site) breathers was established
in~\cite{Bambusi2}. Spectral stability of excited (multi-site) breathers was classified
near the AC limit in the work of~\cite{Archilla,KK09,PelSak}, depending on the phase difference in the nonlinear oscillations
between different sites of the lattice. More recently,
nonlinear instability of spectrally stable two-site breathers was shown
in~\cite{CKP}.  Nevertheless, an overarching criterion of breather
stability tantamount to the VK criterion remains unknown up to now.

In this work, we fill in this important void
by deriving a universal energy criterion {\it both}
for the KG and FPU lattices.
In particular,
we show that a transition from stability  to  instability of a discrete breather
will occur at frequency $\omega$, where the {\it energy-frequency dependence
features an extremum}, i.e., at $H'(\omega)=0$, where $H$ is the breather's energy.
The previously known lattices that exhibit
energy thresholds for discrete breathers like in~\cite{fkm,bernardo}
represent case examples of such an instability transition. Yet, here
we illustrate the generality of such a conclusion both through
an analytical theory and through a number of prototypical numerical
examples (KG, monoatomic FPU, and diatomic FPU).
In the vicinity of the bifurcation point, where $H'(\omega) = 0$,
our asymptotic analysis and
numerical computations suggest the following general conclusion:
Breathers with increasing (decreasing) energy-frequency dependence are generically
unstable in soft (hard) nonlinear potentials. On the other hand, breathers with decreasing (increasing) energy-frequency dependence
in soft (hard) potentials are generally free of the instability associated with this criterion,
yet they may experience other instability forms (including e.g.
period doublings, oscillatory instabilities, etc.~\cite{aubry,FlachPR2008}).
Let us mention that here, the potential is referred to as hard (soft)
when the energy-frequency dependence of individual oscillators
is monotonically increasing (decreasing) \cite{Supp}.\\

\begin{figure}
\begin{tabular}{cc}
\includegraphics[width=4cm]{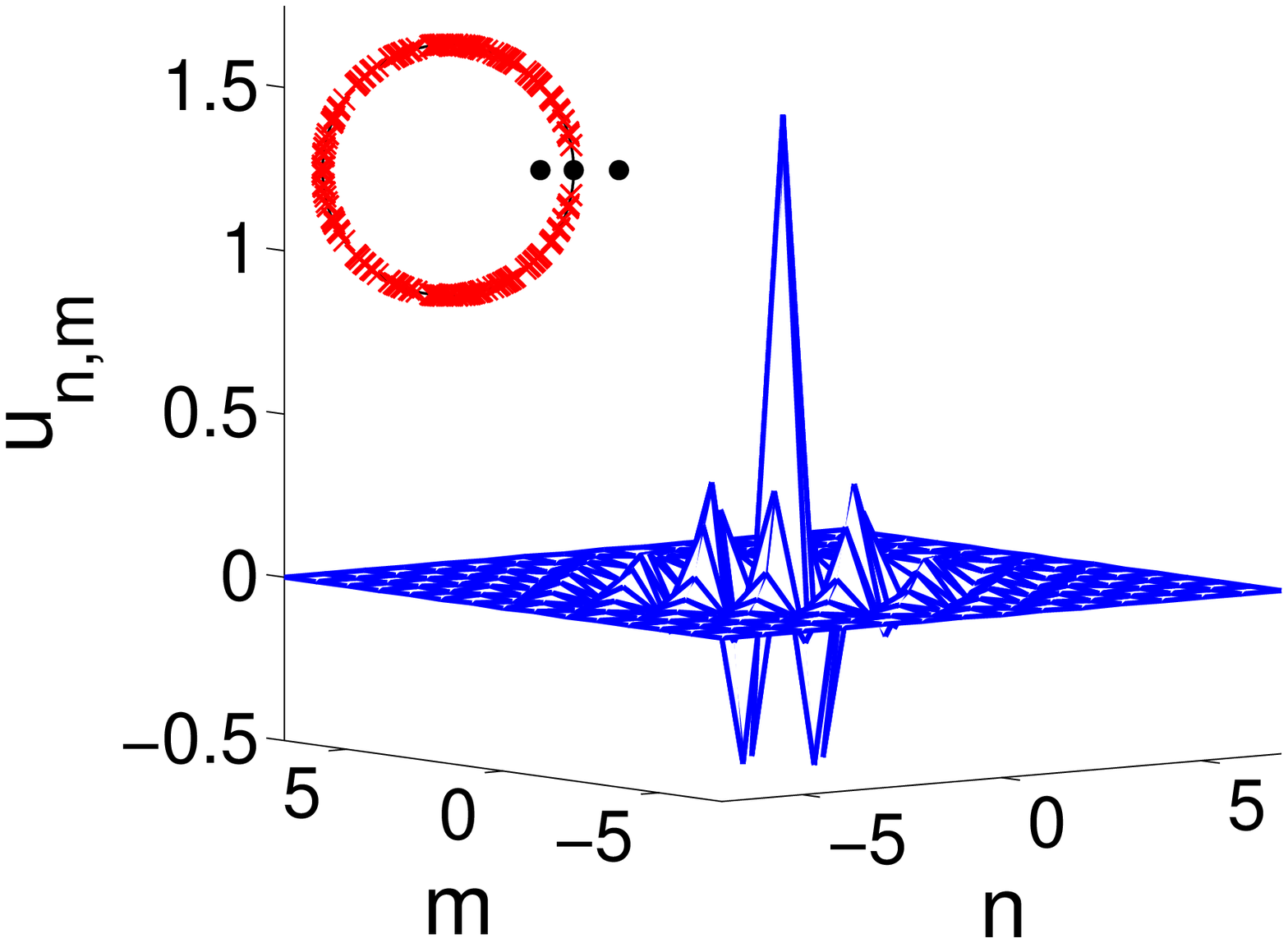} &
\includegraphics[width=4cm]{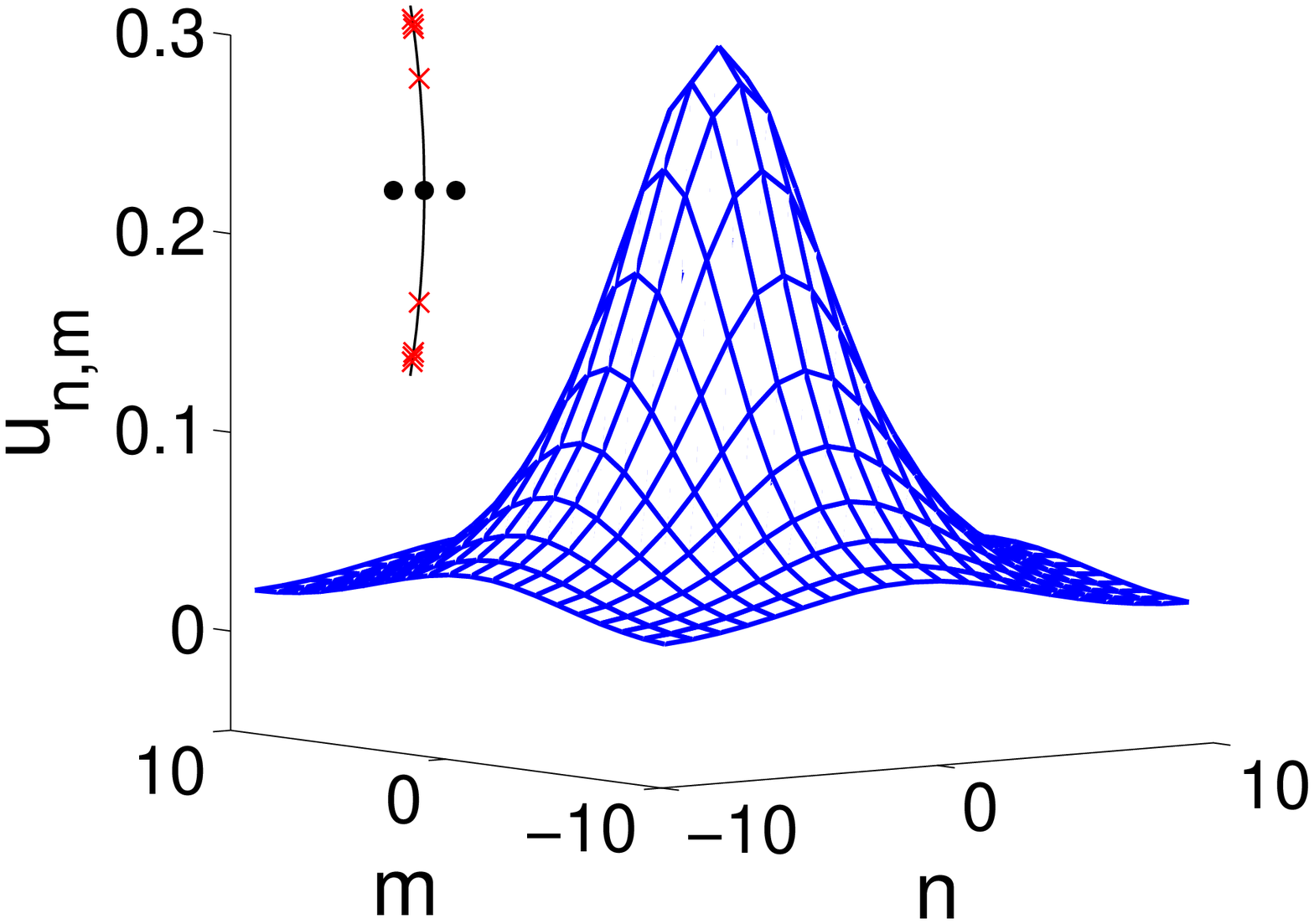} \\
\includegraphics[width=4cm]{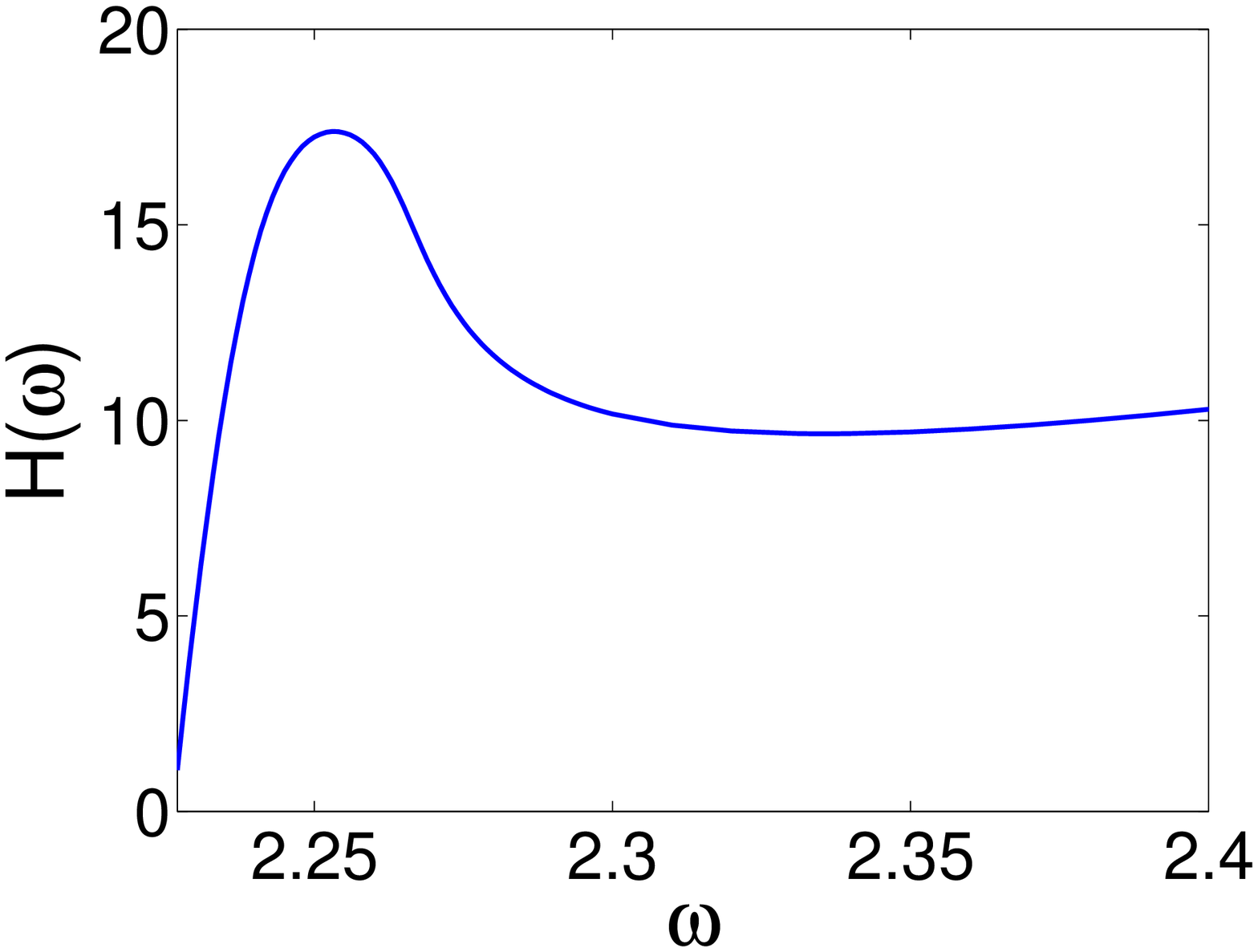} &
\includegraphics[width=4cm]{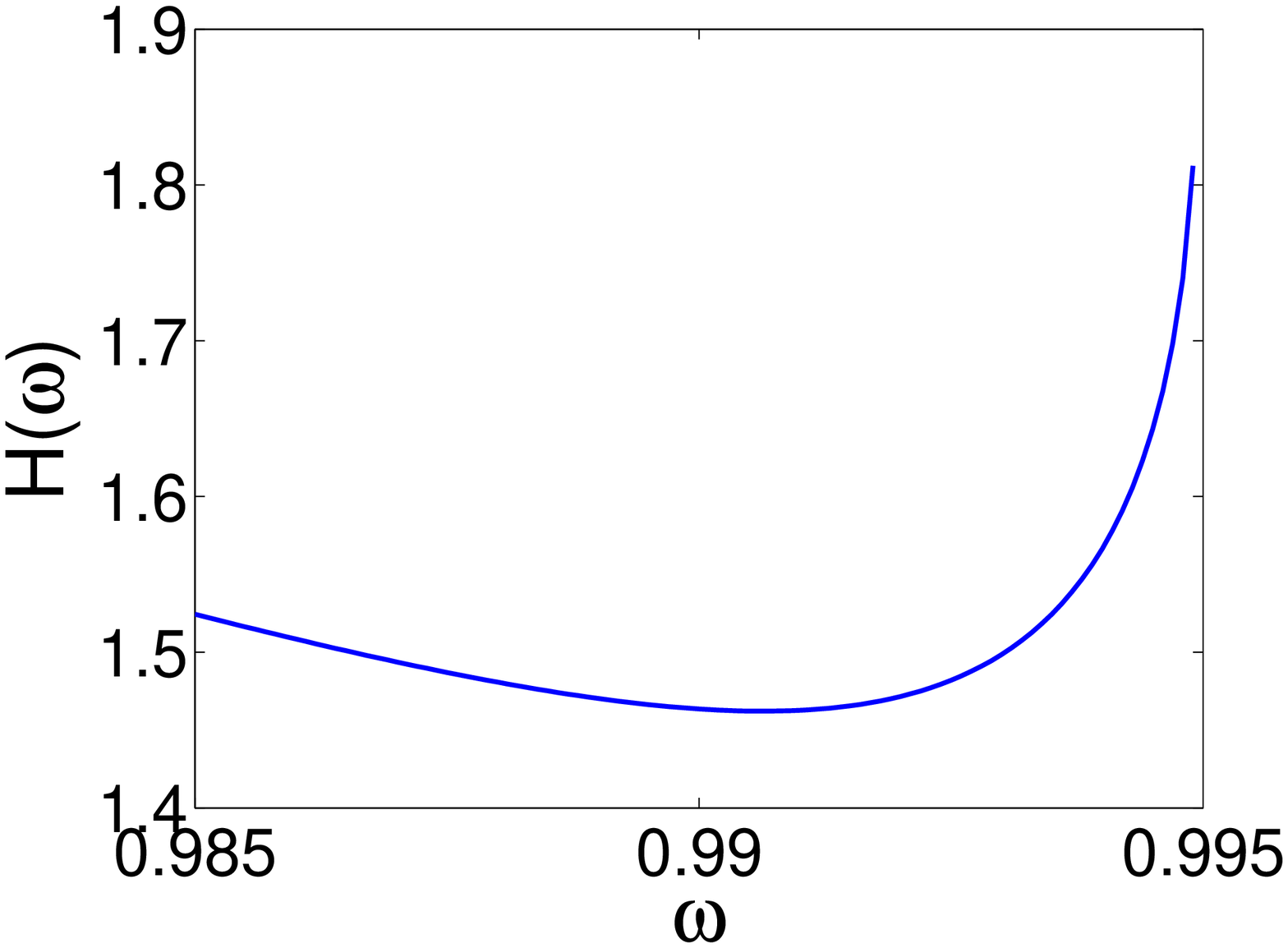} \\
\includegraphics[width=4cm]{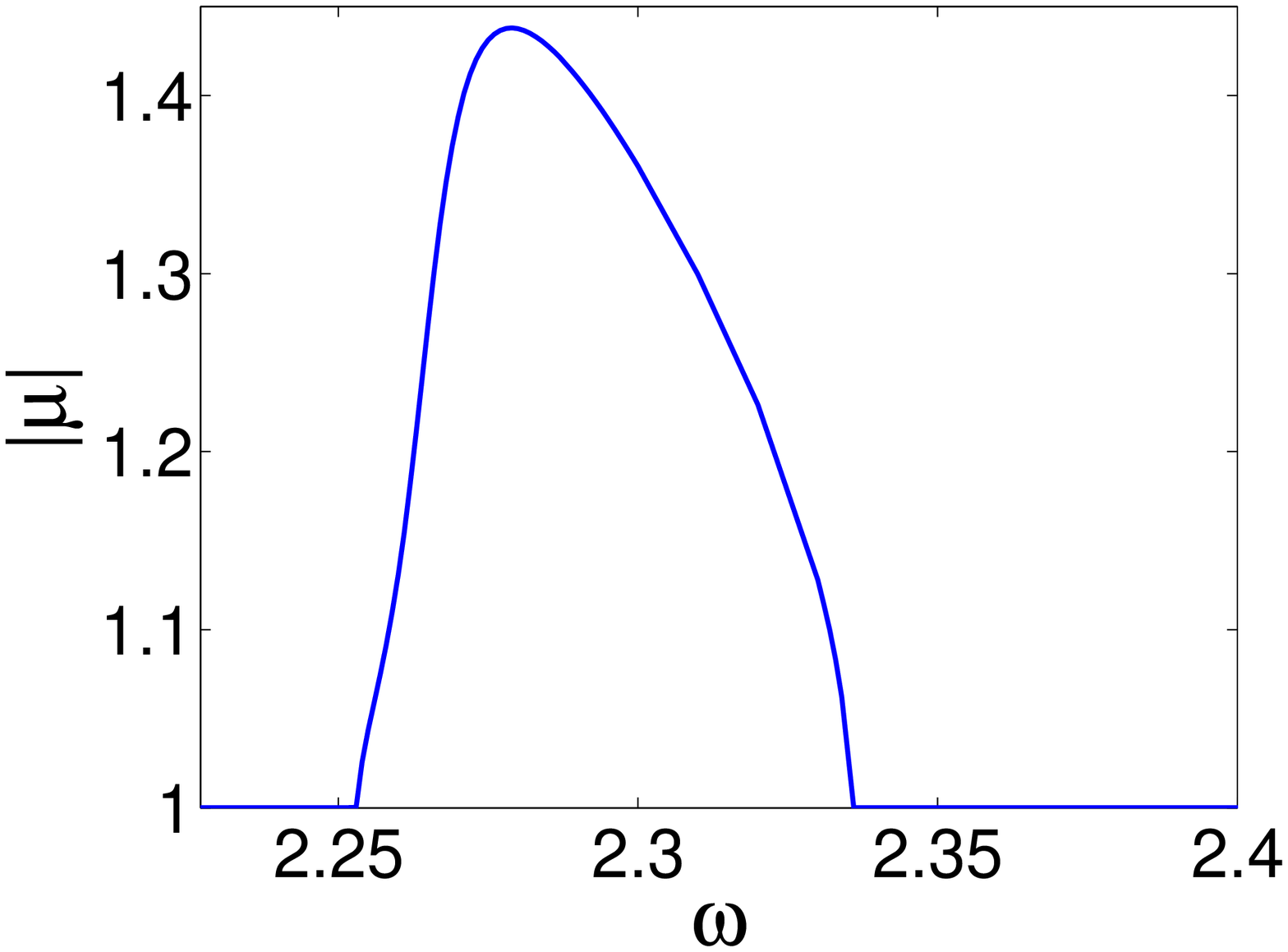} &
\includegraphics[width=4cm]{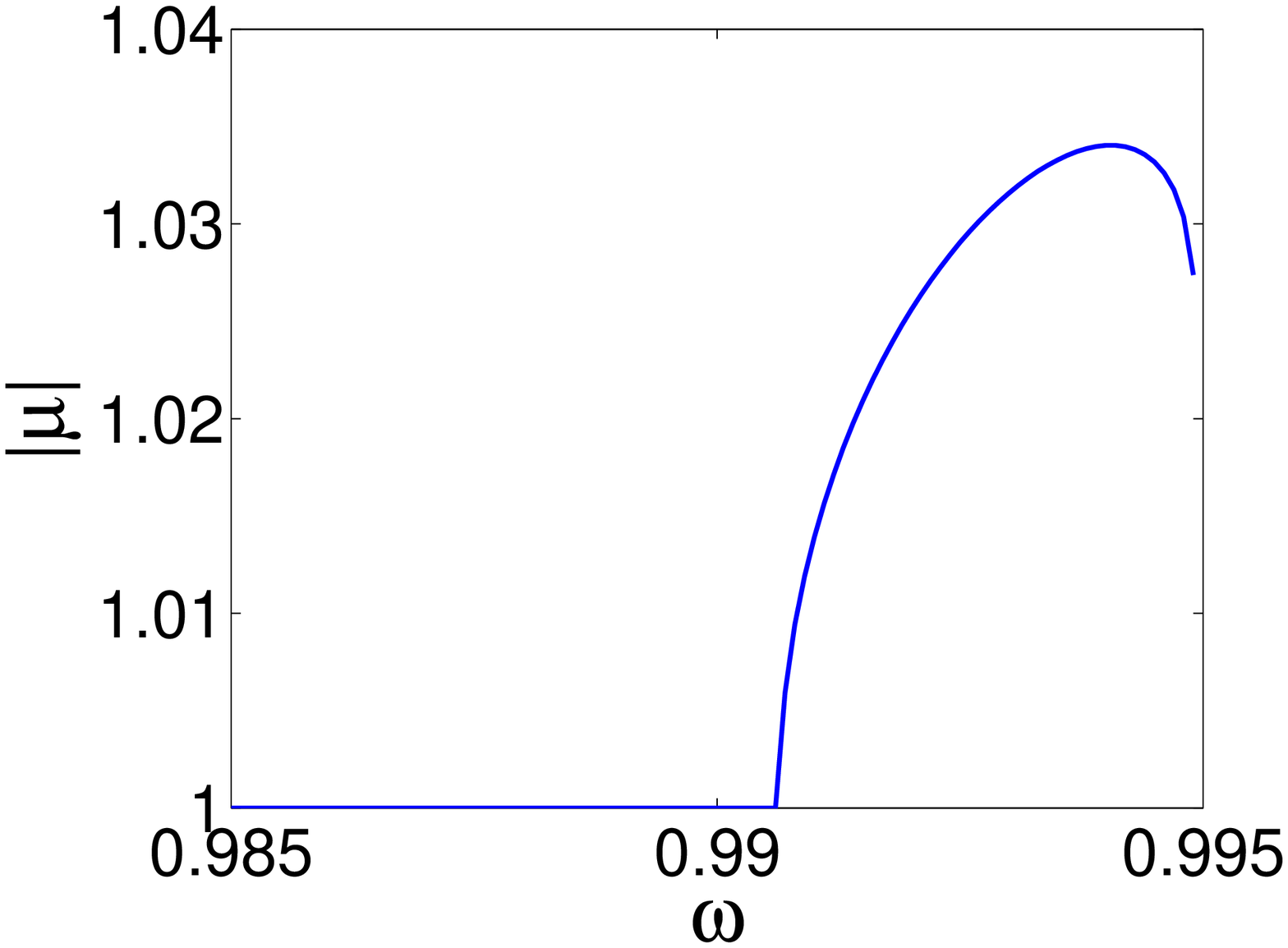} \\
\end{tabular}
\caption{{Breathers in a 2D KG lattice with a hard quartic potential
in the case of $C=0.5$ (left panels) and a Morse potential with $C=0.2$ (right panels).
The top panels show the profile of two unstable breathers with
a portion of the unit circle shown in the inset, corresponding to $C=0.5$, $\wb=2.3$ (left)
and $C=0.2$, $\wb=0.992$ (right). Central panels shows the energy-frequency dependence,
whereas the bottom panels display the Floquet multipliers
with $|\mu|>1$ (i.e., associated with instability) versus $\omega$.}}
\label{fig:KG}
\end{figure}

{\it Mathematical Setup.}  We consider a one-dimensional (1D) chain
of nonlinear oscillators under Newtonian dynamics:
\begin{eqnarray}\label{lattice}
    \ddot u_n+V'(u_n) = W'(u_{n+1}-u_n) - W'(u_n-u_{n-1}),
\end{eqnarray}
where $n$ is defined on a 1D lattice, $V$ is an on-site (substrate) potential and $W$ is
the inter-site potential for nearest-neighbor interaction. Both $V$ and $W$ are assumed smooth.
The associated energy function for the lattice (\ref{lattice}) is given by
\begin{eqnarray}\label{energy}
H = \sum_{n \in \mathbb{Z}} \frac{1}{2} \dot{u}_n^2 + V(u_n) + W(u_{n+1}-u_n).
\end{eqnarray}
If $W'(u) = C u$ with coupling constant $C$ while $V$ satisfies $V'(0) = 0$ and $V''(0) > 0$, the chain is referred to as the
Klein--Gordon (KG) lattice. If $V'(u) = 0$ while $W$ satisfies $W'(0) = 0$ and $W''(0) > 0$,
the chain is referred to as the Fermi--Pasta--Ulam (FPU) lattice.
For clarity, we describe our results for the KG lattice and draw parallels
to the FPU case.

Discrete breathers of the KG lattice are $T$-periodic solutions
with $u_n(t + T) = u_n(t)$ for every $n$.
Setting the breather frequency to $\omega = 2\pi/T$, we can
normalize the period of the breather to $2\pi$
using $u_n(t) = U_n(\tau)$, where $\tau = \omega t$
and $U_n(\tau+2\pi) = U_n(\tau)$. The profile $U_n$ also depends
on frequency $\omega$. We then have
\begin{eqnarray}\label{breather}
    \omega^2 U_n''(\tau) + V'(U_n(\tau)) = C (\Delta U)_n(\tau),
\end{eqnarray}
where $(\Delta U)_n$ denotes the discrete Laplacian.
The spectral stability of discrete breathers is determined
by the linearized equations of motion
\begin{equation}
\label{linKG}
\ddot w_n + V''(u_n) w_n = C (\Delta w)_n,
\end{equation}
where $w_n$ is a perturbation to $u_n$. According to the Floquet theory, we are looking for solutions
of the linearized equation (\ref{linKG}) in the form $w_n(t) = e^{\lambda t} W_n(\tau)$,
where $\lambda \in \mathbb{C}$ is a spectral parameter and $W_n(\tau+2\pi) = W_n(\tau)$. The spectral stability
problem is then
\begin{eqnarray}
\label{spectrumKG}
\omega^2 W_n''(\tau) &+& 2 \lambda \omega W_n'(\tau) + \lambda^2 W_n(\tau)
\nonumber
\\
& + &  V''(U_n(\tau)) W_n(\tau) = C (\Delta W)_n(\tau).
\end{eqnarray}

The (continuous) spectral bands can be identified on the unit circle
in terms of the Floquet multipliers $\mu = e^{\lambda T}$.
To be precise, the two bands are located at $\mu_{\pm}(\theta) = e^{\pm i \omega(\theta) T}$,
where $\omega(\theta) = \sqrt{1+4C \sin^2\left(\frac{\theta}{2}\right)}$,
$\theta \in [-\pi,\pi]$.
We assume that the two bands are bounded away from the unit multiplier $\mu_0 = 1$,
which corresponds to the isolated eigenvalue $\lambda_0 = 0$ in the spectral problem (\ref{spectrumKG}).
Because of the translational invariance
symmetry (in time), we note that the isolated
eigenvalue $\lambda_0 = 0$ is at least double. Indeed, the
eigenvector $W_n(\tau) = U_n'(\tau)$ satisfies (\ref{spectrumKG}) for $\lambda = 0$.
Furthermore, the generalized eigenvector $\tilde{W}_n(\tau) = \partial_{\omega} U_n(\tau)$
satisfies the derivative of (\ref{spectrumKG}) in $\lambda$ for $\lambda = 0$
given by
\begin{equation}
\label{linear-0}
(L \partial_{\omega} U)_n(\tau) = 2 \omega U_n''(\tau),
\end{equation}
where
$$
(L W)_n(\tau) = C (\Delta W)_n(\tau) - V''(U_n(\tau)) W_n(\tau) - \omega^2 W_n''(\tau)
$$
is the linearized operator for the spectral problem (\ref{spectrumKG}).

Let us assume that the kernel of $L$ is exactly
one-dimensional with the eigenvector $W_n(\tau) = U_n'(\tau)$. This assumption is generally
satisfied because no other symmetry exists in the lattice (\ref{lattice}) besides the
translational symmetry in time.
The most typical scenario of a discrete breather becoming unstable occurs when
a pair of Floquet multipliers $\mu$ on the unit circle coalesces at
$\mu_0 = 1$ and splits along the real
axis. At the critical point,  the eigenvalue $\lambda_0 = 0$ of the spectral problem (\ref{spectrumKG})
is assumed to have a higher-than-two-algebraic multiplicity.
It is exactly that condition which will provide us with the
energy criterion for spectral stability of discrete breathers, as follows.

\begin{figure}[h!]
\begin{tabular}{cc}
\includegraphics[width=4cm]{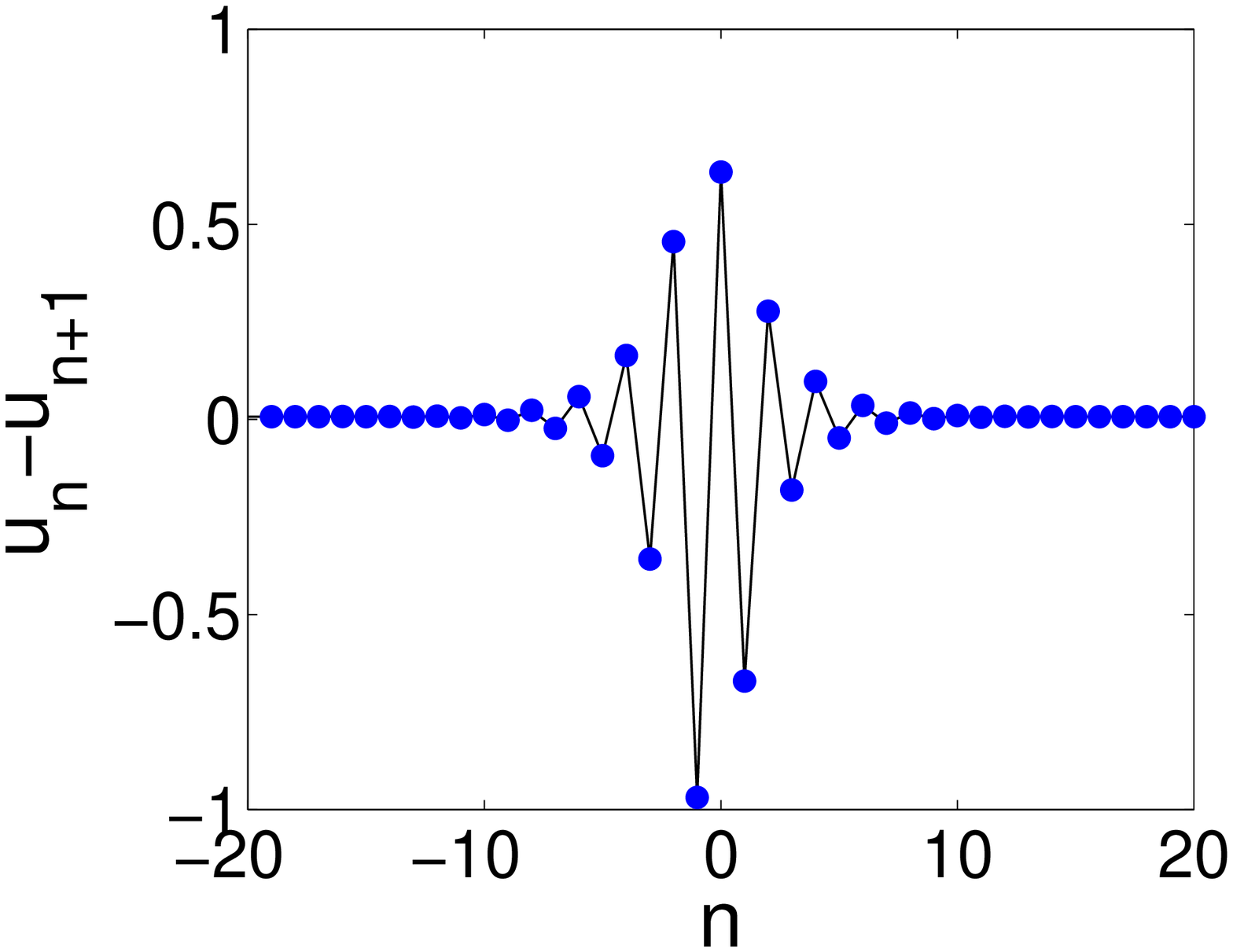} &
\includegraphics[width=4cm]{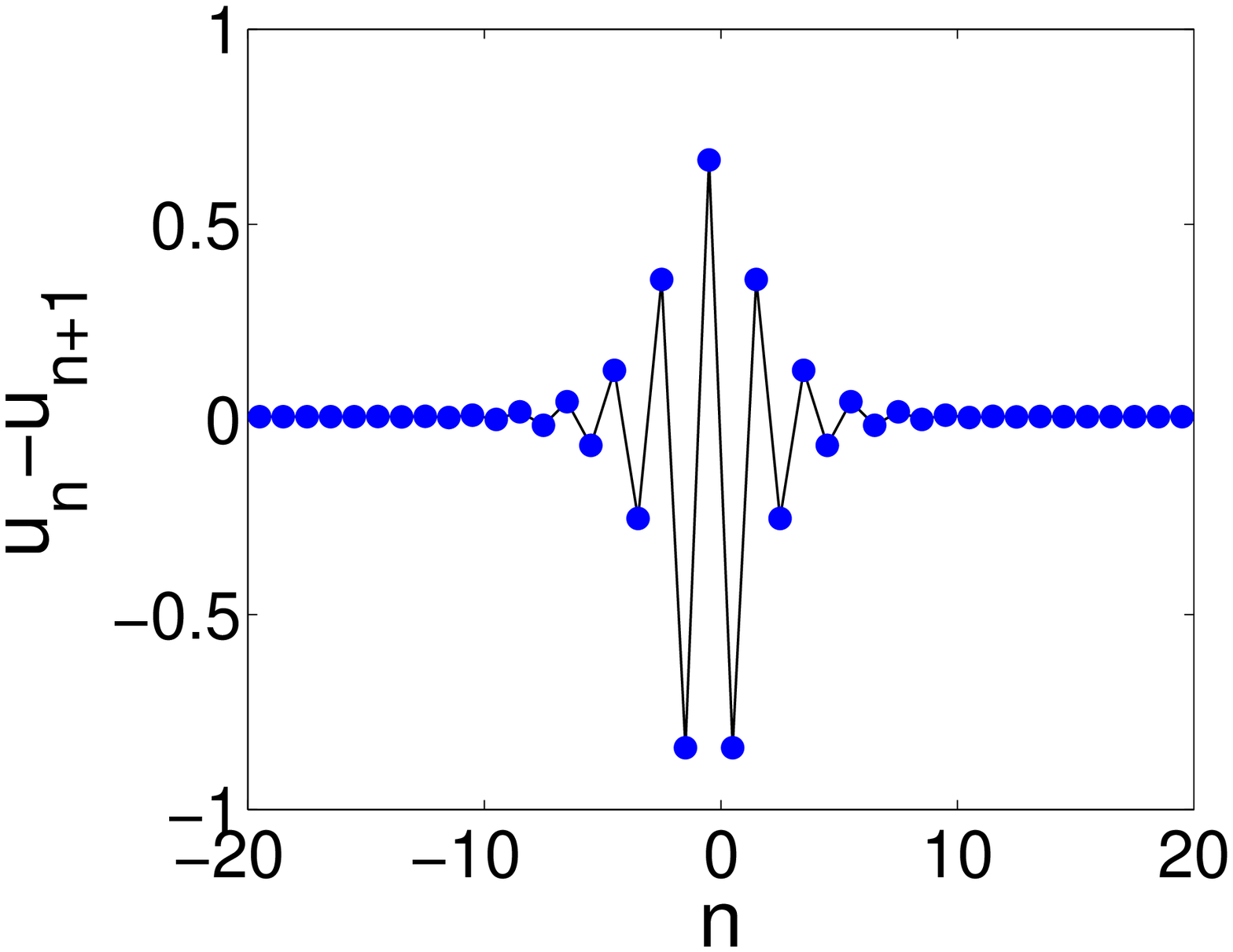} \\
\includegraphics[width=4cm]{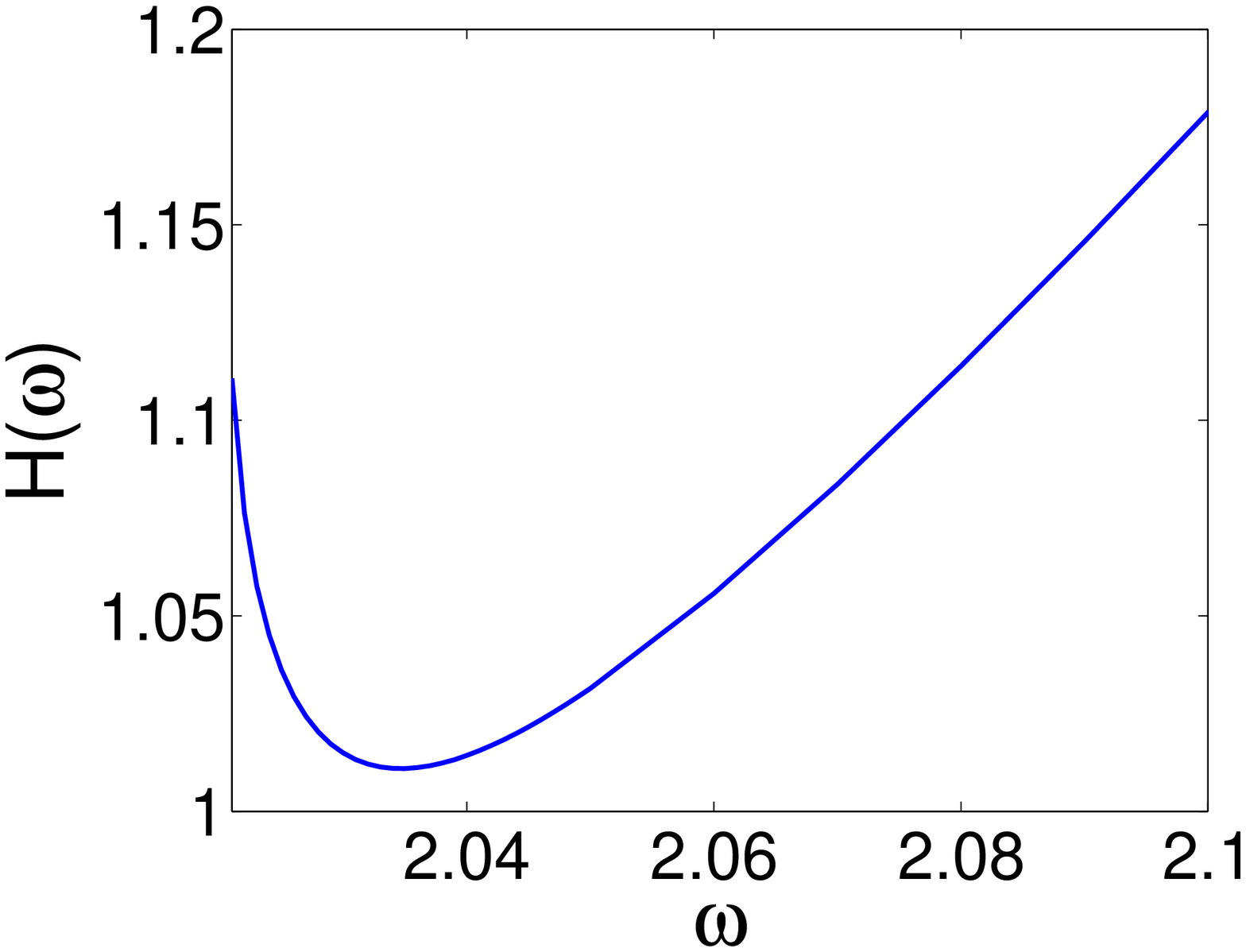} &
\includegraphics[width=4cm]{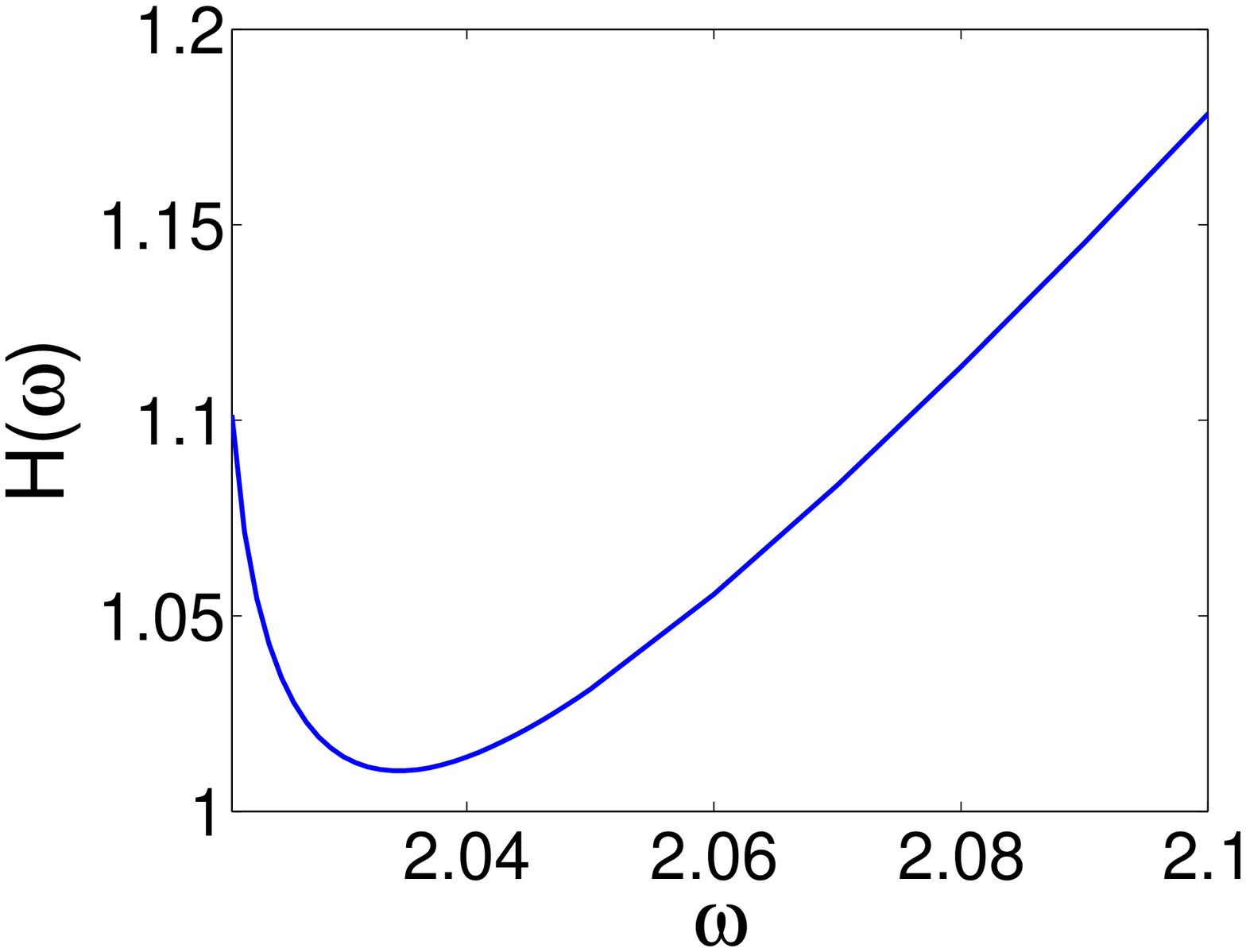} \\
\includegraphics[width=4cm]{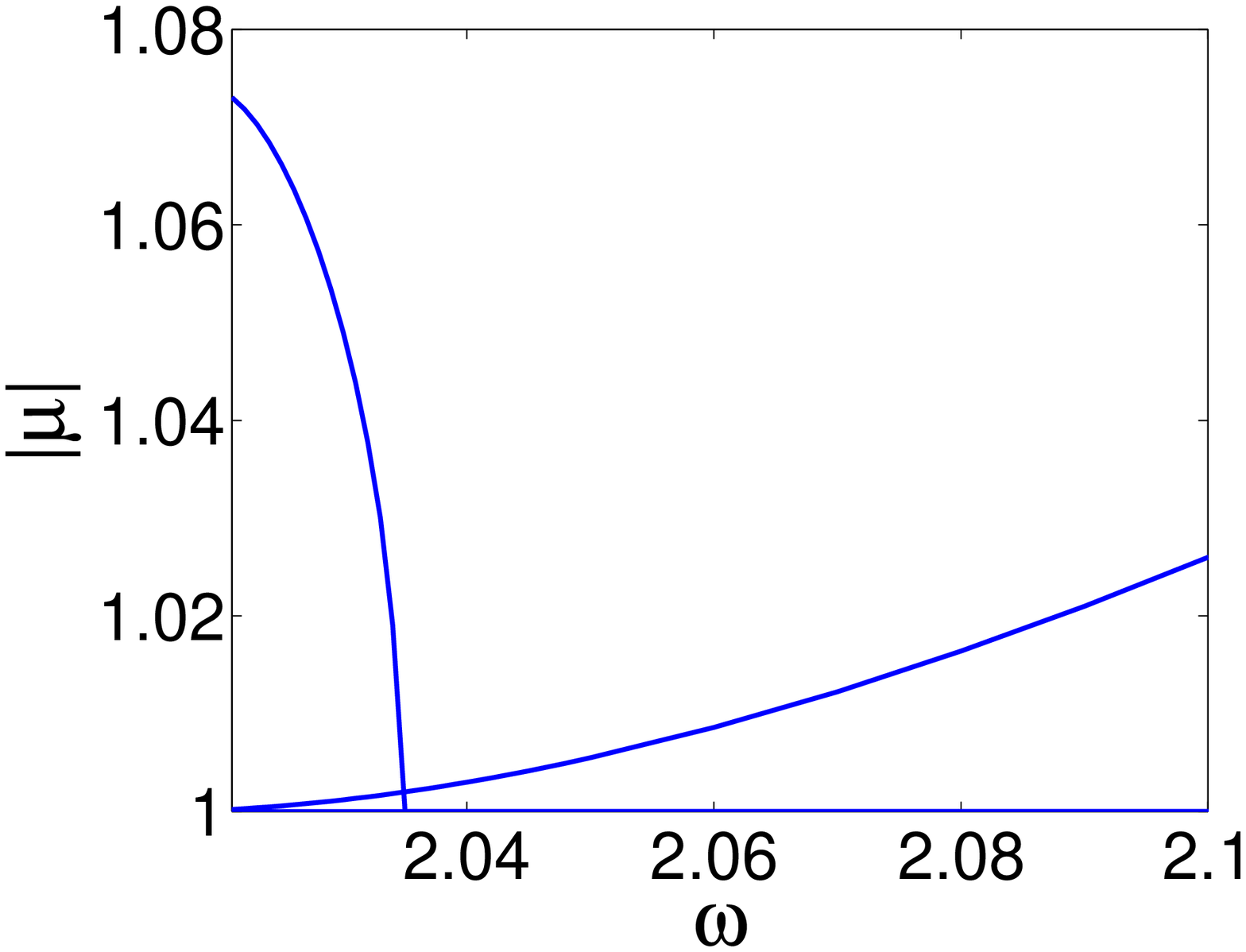} &
\includegraphics[width=4cm]{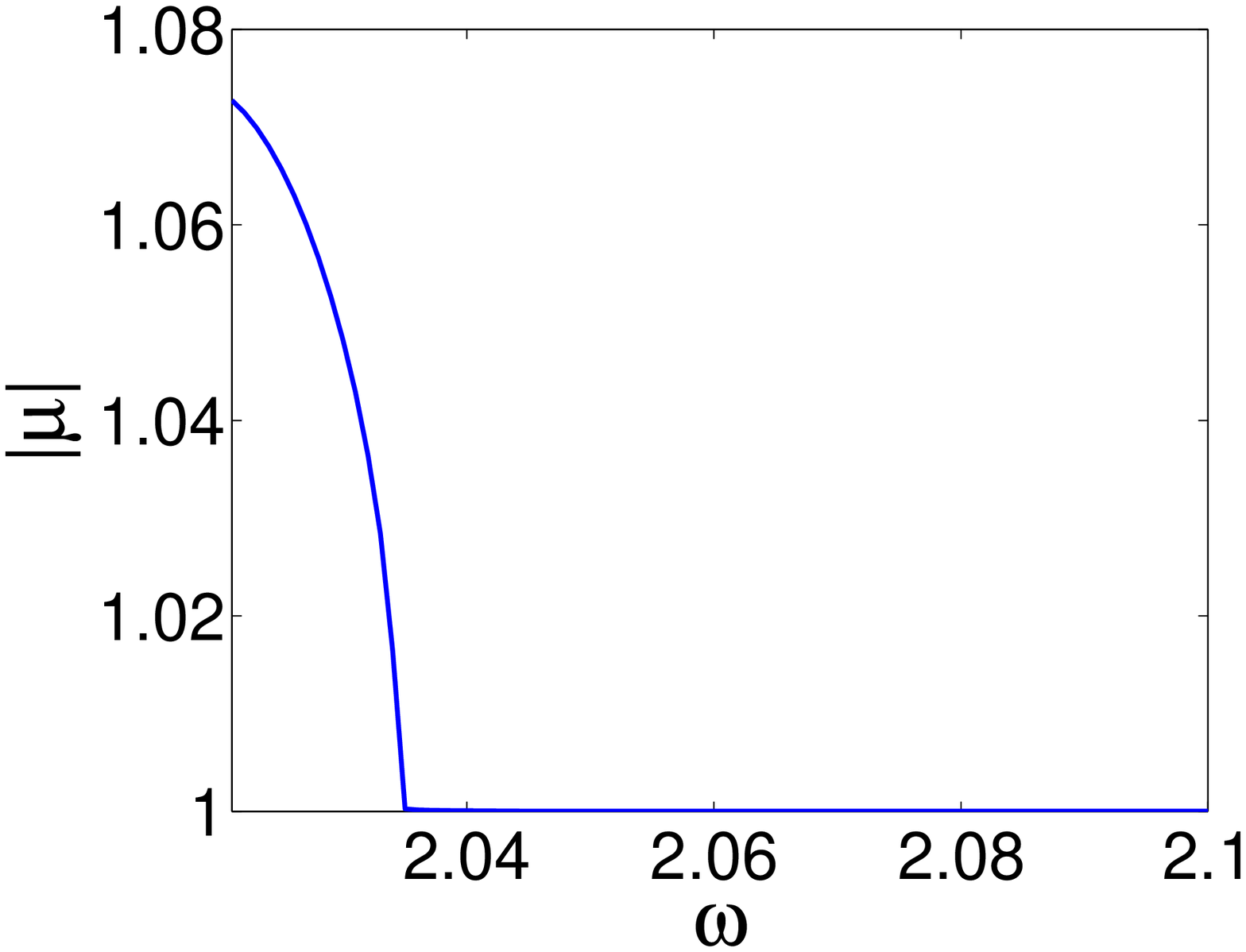} \\
\end{tabular}
\caption{Breathers in a monoatomic FPU chain with $\alpha=-1$, $\beta=1$. Left (right)
panels corresponds to the Sievers--Takeno (Page) mode. The top panels show the breather profiles, in the strain
variable, for $\wb=2.1$. The middle panel shows the energy-frequency dependence,
whereas the bottom panel displays modulus of the Floquet multipliers with $|\mu|>1$ versus $\omega$.}
\label{fig:FPUmono}
\end{figure}

\begin{figure}[h!]
\begin{tabular}{cc}
\includegraphics[width=4cm]{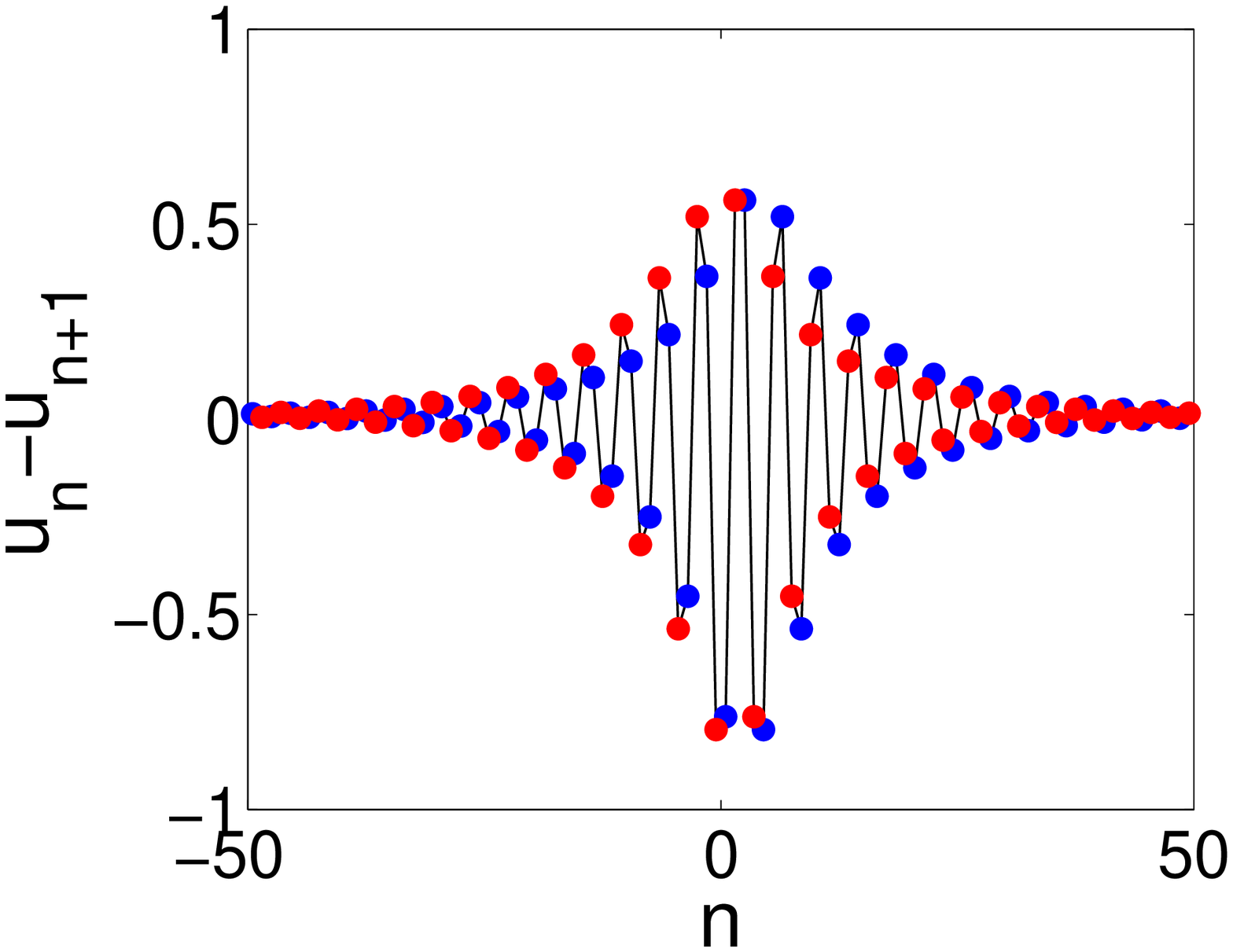} &
\includegraphics[width=4cm]{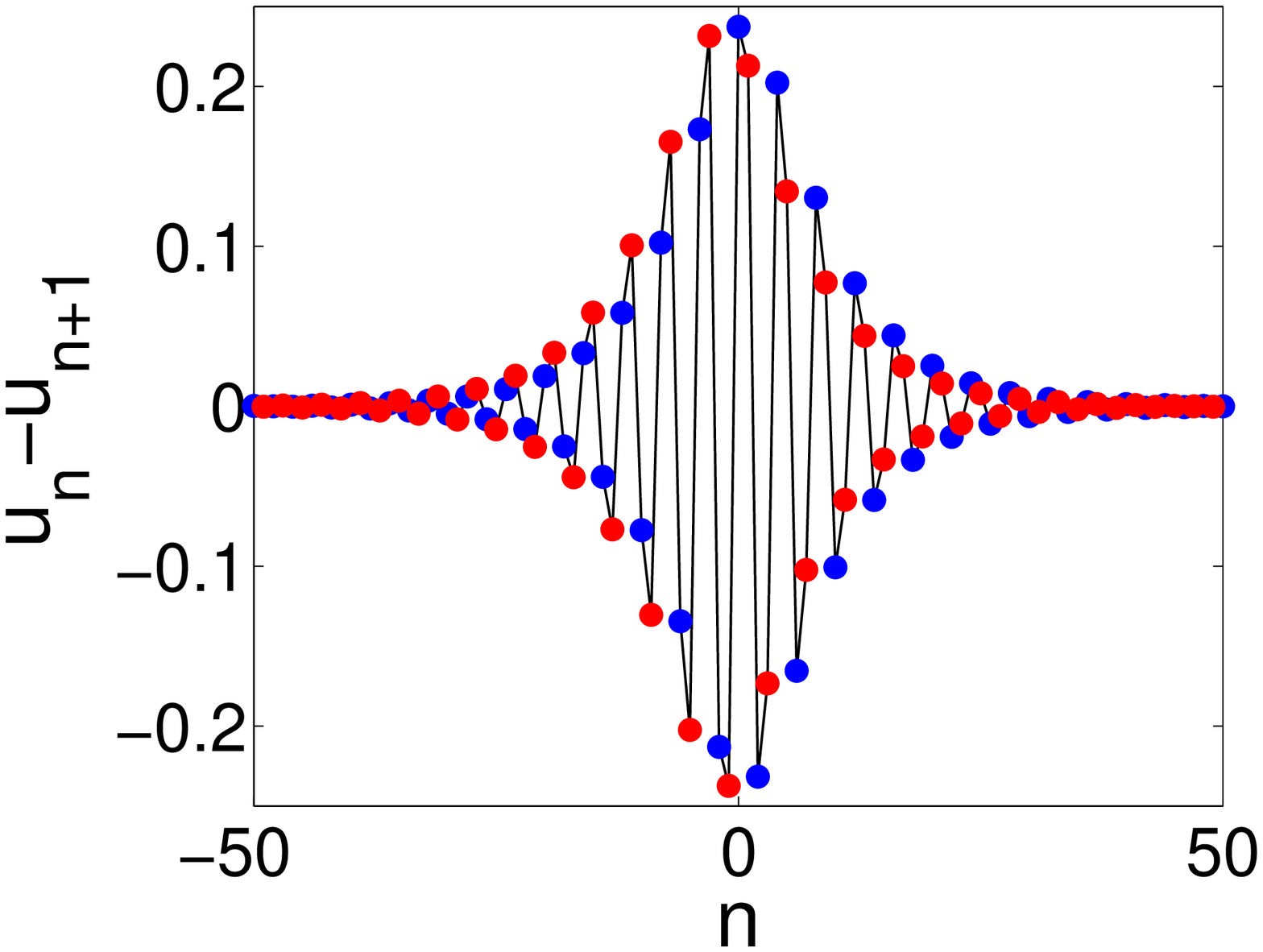} \\
\includegraphics[width=4cm]{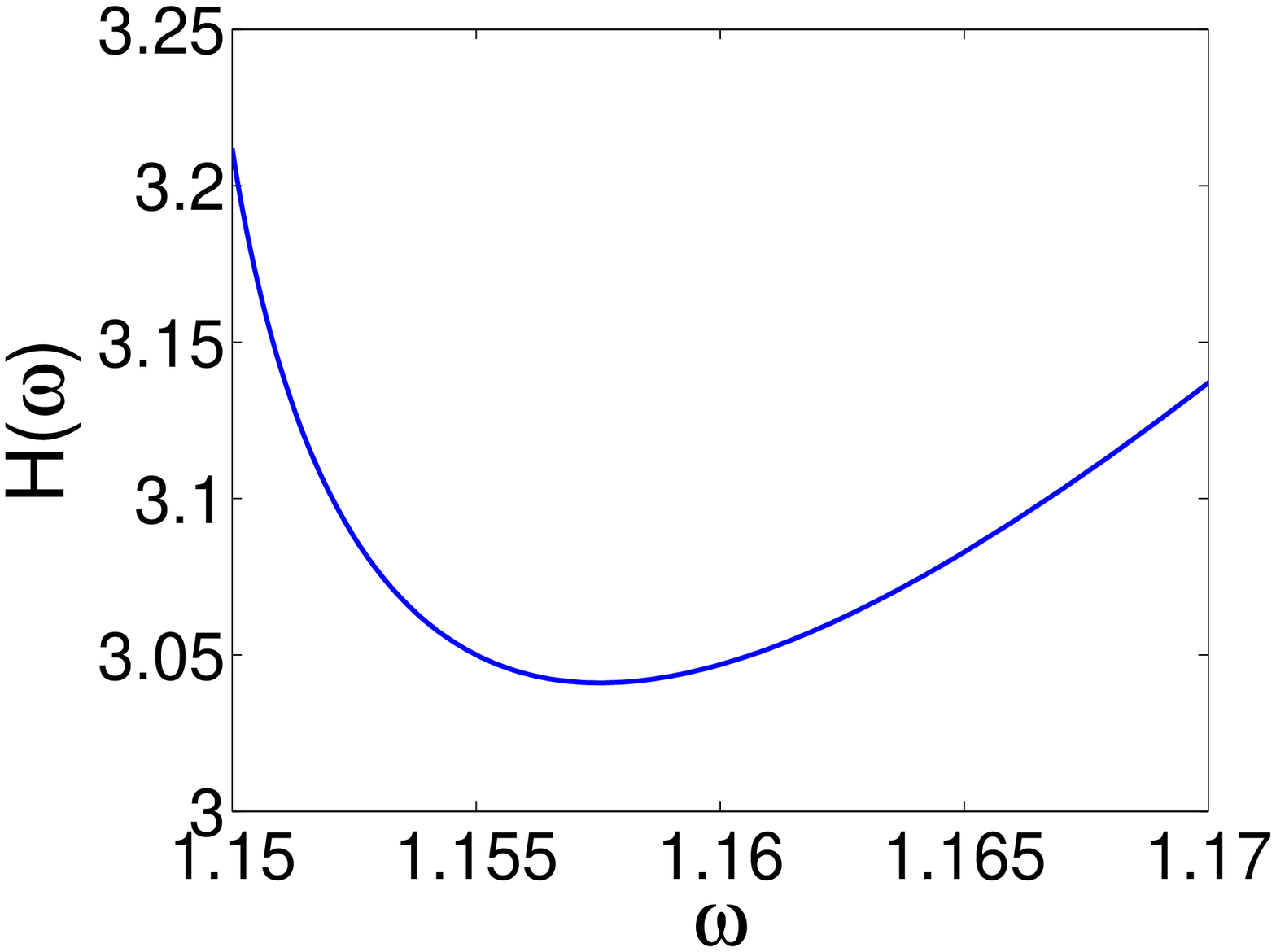} &
\includegraphics[width=4cm]{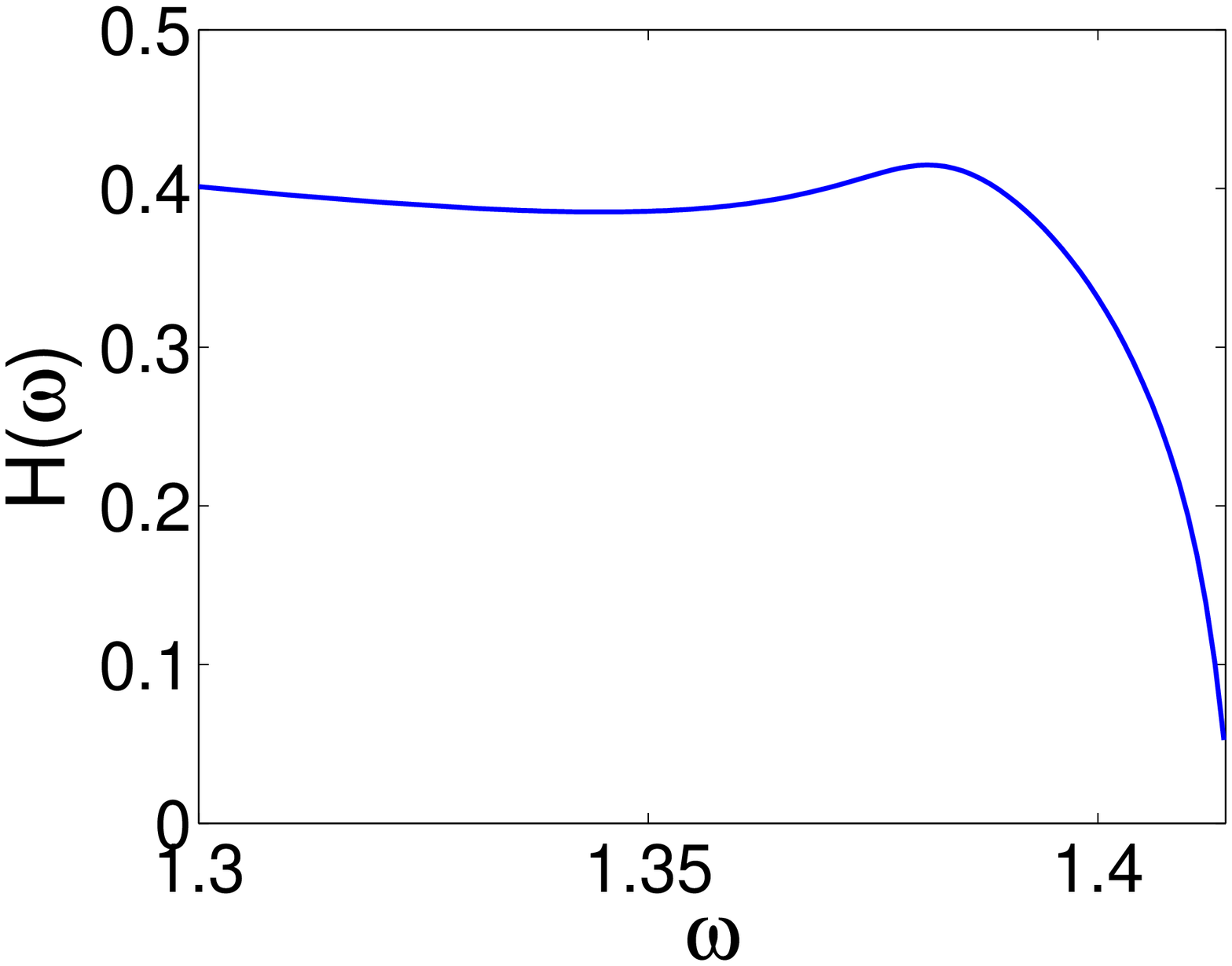} \\
\includegraphics[width=4cm]{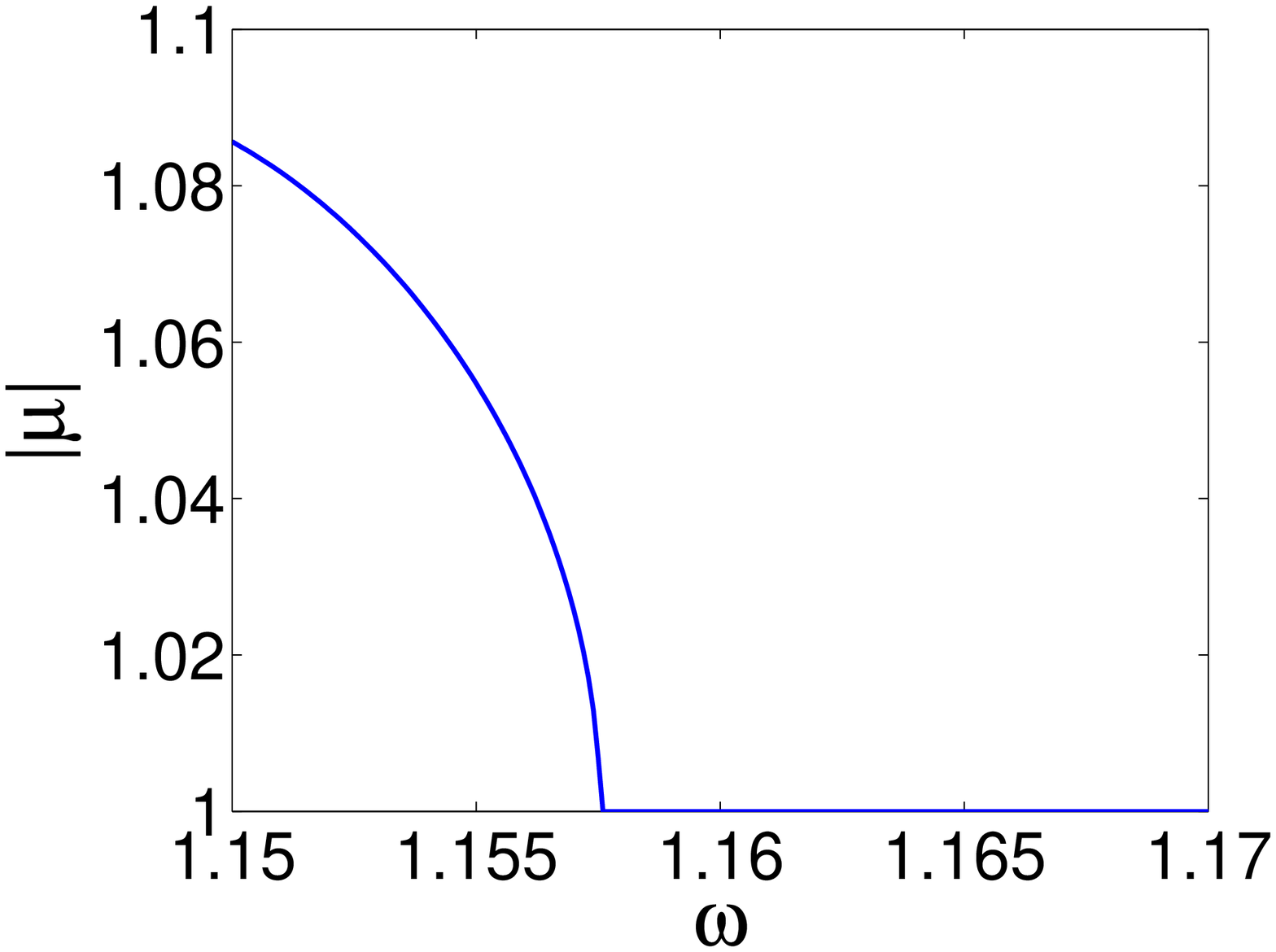} &
\includegraphics[width=4cm]{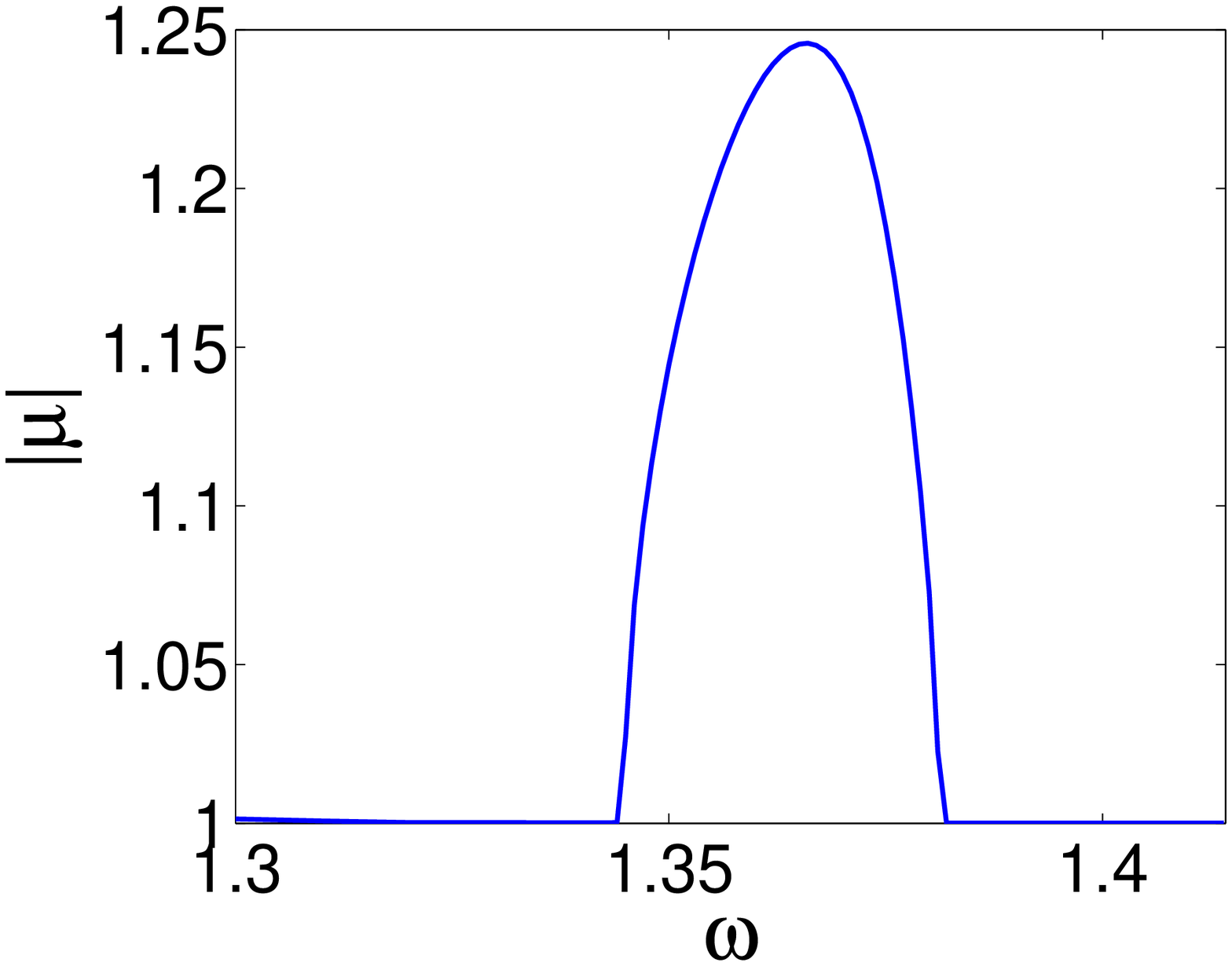} \\
\end{tabular}
\caption{{Gap breathers in a diatomic FPU chain for a {\em hard} potential
with $\alpha=-1$, $\beta=1$, $\epsilon=0.8$ (left) and a {\em soft} potential
with $\alpha=0$, $\beta=-1$ and $\epsilon=0.7$ (right).
The top panels show the breather profiles, in the strain variable, for  $\wb=1.7$ (left) and $\wb=1.4$ (right).
Blue (red) dots correspond to the more (less) massive particles. The middle panel shows the
energy-frequency dependence, whereas the bottom panel displays the modulus of the Floquet multipliers
with $|\mu|>1$ versus $\omega$.}}
\label{fig:FPUdiat}
\end{figure}

The condition that $\lambda_0 = 0$ is at least quadruple (by Hamiltonian symmetry, it has an even multiplicity)
is equivalent to the Fredholm condition of existence of a solution to the second derivative of (\ref{spectrumKG})
in $\lambda$ for $\lambda = 0$. Using the projection technique \cite{Supp}
yields the solvability condition in the form
\begin{eqnarray*}
0 = \int_{0}^{2\pi} \sum_{n \in \mathbb{Z}} U_n'(\tau) \left[ 2 \omega \partial_{\omega} U_n'(\tau) + U_n'(\tau) \right] d \tau = T H'(\omega),
\end{eqnarray*}
where $H(\omega)$ is the time-independent breather energy that follows from (\ref{energy}).
The higher multiplicity condition (signaling the potential
transition between stability and instability)
is thus
satisfied if $\omega$ is a critical point of the breather energy $H(\omega)$.

The solvability condition $H'(\omega) = 0$ cannot be satisfied in the AC limit,
where the individual oscillator is always stable with
$H'(\omega) > 0$ for hard potentials and $H'(\omega) < 0$ for soft potentials \cite{Supp}.
However, far from the AC limit such a bifurcation may
(and often does) occur. If
at the critical point, $\lambda_0=0$ is exactly quadruple,
i.e., if a pair of simple Floquet multipliers coalesces
with the double unit multiplier $\mu_0 = 1$ at $H'(\omega) = 0$,
then an expansion of the eigenvalue problem (\ref{spectrumKG})
near the bifurcation point yields:
\begin{equation}
\label{dispersion}
\lambda^2 T H'(\omega) + \lambda^4 M + \mathcal{O}(\lambda^6) = 0,
\end{equation}
where $M \neq 0$. Then, if $M > 0$, the breathers
are stable if $H'(\omega) > 0$ and unstable if $H'(\omega) < 0$, whereas if $M < 0$,
then the breathers are stable if $H'(\omega) < 0$ and unstable if $H'(\omega) > 0$.
Detailed asymptotic analysis \cite{Supp} suggests that
the former case is intrinsic for hard potentials and the latter
case is typical for soft potentials, at least in the small-amplitude limit of KG breathers.

The same conclusion is also drawn in the FPU case when reformulated
in terms of the strain variable $r_n=u_{n+1} - u_n$, because it is the strain variable
that decays to zero at infinity for FPU breathers \cite{Supp}. \\

{\it Numerical illustrations: 2D KG breathers.} We consider a two-dimensional (2D)
version of the KG lattice with the hard $\phi^4$ potential $V(u)=u^2/2+u^4/4$~\cite{SatoC2003}
and the soft Morse potential $V(u)=(\exp(-u)-1)^2/2$. The latter has
been ubiquitously utilized for the study of breathers
in DNA denaturation settings where it is used
to model the hydrogen bond connecting the two bases in a
pair~\cite{PeyrardBishop}.

Fig.~\ref{fig:KG} shows the energy-frequency dependence
for a fixed coupling constant {$C$}, as well as the most unstable real
Floquet multiplier (recall that instability is tantamount to $|\mu|>1$)
{for both hard and soft potentials}.
We observe a perfect correlation, as prescribed by the theory,
between the stability changes and energy extrema. Indeed, the
breather is stable (unstable) at the regions of increasing
(decreasing) energy $H(\wb)$ for hard potentials, and this trend is reversed for soft potentials.

Notice that in the case of the hard potential, the breather is still stable for every $\wb$ past the upper
limit shown in Fig.~\ref{fig:KG}. However, in the case of the Morse potential, an
instability emerges for $\wb$ below
the lower limit of the figure. This instability not predicted by our energy criterion pertains to
the exchange of instability (precursor of breather mobility) that typically occurs
within the Morse potential \cite{cretegny}. \\

{\it Numerical Illustrations: 1D FPU breathers.} We consider both monoatomic and diatomic FPU
chains~\cite{Chaos}. In general, these chains are modeled by the FPU equation
\begin{equation}
  M_n \ddot{u}_n = W'(u_{n+1}-u_n) - W'(u_n-u_{n-1}),
\end{equation}
with $M_n$ being the particle masses. We choose $V(u)=u^2/2+\alpha u^3/3+\beta u^4/4$.
In the monoatomic case, $M_n=1$ for all sites, whereas, in the diatomic case,
$M_n=1$ for $n$ even and $M_n=1/\epsilon^2$ for $n$ odd,
where $\epsilon^2$ is the parameter for mass ratio of the diatomic FPU chain~{\cite{cls,Noble}}.

It was demonstrated in \cite{bernardo}, for the {\em monoatomic chain}, that
the large-amplitude breathers possess a minimum of $H(\wb)$ since
their amplitude does not tend to zero at the band edge $\wb \to 2$.
The energy threshold exists when $\alpha$ is taken below a critical
value of $\alpha_c = -\sqrt{3}/2 \approx -0.86$ (for $\beta=1$).
However, in \cite{bernardo}, the instability past the energy minimum was not considered.
Here we show that the energy threshold results in the change of stability of discrete breathers.

As is typically the case in both FPU and KG chains, there are two principal breathers,
the so called Sievers--Takeno (bond-centered) and Page (site-centered) modes. The former is, in general,
exponentially unstable. Fig. \ref{fig:FPUmono} shows, as dictated by our
stability criterion for hard potentials, that an exponential instability arises
at the energy minimum for {\em both} modes when $\wb \to 2$.
In the Page mode, this transition manifests itself as the appearance of
an exponential instability of the previously stable structure.
In the already unstable Sievers--Takeno mode, a second unstable Floquet multiplier appears
as $\wb \to 2$ (for a secondary instability which
rapidly overtakes the previous one as the instability with the
largest growth rates).

In the {\em diatomic case}, there is an opening of a frequency gap within the phonon spectrum,
$$
2 \epsilon^2 W''(0) < \omega^2 < 2 W''(0).
$$
This allows the existence of breathers with frequency $\omega$
in the gap of the phonon spectrum (so-called {\em gap breathers}).
Such structures can exist even in the case of soft potentials~\cite{Kastner},
bifurcating from the bottom of the optical phonon band; see
also~\cite{BoechlerPRL2010} for a relevant experimental manifestation of
such modes. For the soft potential, see the right panels on Fig. \ref{fig:FPUdiat},
no global energy minimum exists but extrema in the energy-frequency curve may occur even if $\alpha = 0$.
In a full agreement with the energy criterion for soft potentials,
the instability of such gap breathers is perfectly
correlated with the increasing energy-frequency dependence,

Finally, gap breathers also exist for hard potentials, bifurcating from the
top of the acoustic band, see the left panels on Fig. \ref{fig:FPUdiat}.
Their stability and energetic properties are similar to the breathers in the monoatomic FPU lattice,
also necessitating a non-zero $\alpha$ for the existence of energy minima.\\

{\it Conclusions.} In this work we have presented a
systematic and general energy criterion for spectral stability
of breathers in nonlinear dynamical lattices.
The energy stability criterion for discrete breathers is strongly reminiscent of
the VK criterion for solitary waves; in fact, as illustrated
in \cite{Supp}, it {\it reduces} to the VK criterion in the small
amplitude limit where the breathers can be approximated as
solitary waves. In view of that, the proposed criterion can be
considered as the definitive analogue of the VK criterion for breathers.

We have then corroborated the validity of the energy criterion for stability of discrete breathers
via a wide range of models, both KG and FPU, both 1D
and 2D, both homogeneous and heterogeneous, showcasing that its
generality transcends the specific such properties of the model.
It follows from our numerical results that the breathers
are unstable in hard (soft) potentials if the energy-frequency dependence
is decreasing (increasing) and stable otherwise.

Admittedly, a general classification of instabilities
of breathers (more generally of periodic orbits, including
non-localized ones, such as plane waves in Hamiltonian systems)
in the same spirit as the well developed theory
of solitary waves of the nonlinear Schr{\"o}dinger equation
is still incomplete. Nevertheless, the present criterion
we believe, constitutes an important step towards
future work in this direction, and on understanding nonlinear stability of
breathers in lattices.


\begin{thebibliography}{99}

\bibitem{aubry} S. Aubry, Physica D {\bf 103}, 201 (1997).

\bibitem{FlachPR2008} S.~Flach and A.V.~Gorbach, Phys. Rep. {\bf 467}, 1 (2008).

\bibitem{TriasPRL2000} E.~Tr\'{i}as, J.J.~Mazo and T.P.~Orlando, Phys. Rev. Lett. {\bf 84}, 741 (2000).

\bibitem{BinderPRL2000} P.~Binder, D.~Abraimov, A.V.~Ustinov, S.~Flach and Y.~Zolotaryuk, Phys. Rev. Lett. {\bf 84}, 745 (2000).

\bibitem{SatoPRL2003} M.~Sato, B.E.~Hubbard,  A.J.~Sievers, B.~Ilic, D.A.~Czaplewski and H.G.~Craighead, Phys. Rev. Lett. {\bf 90}, 044102 (2003).

\bibitem{SatoC2003} M.~Sato, B.E.~Hubbard, L.Q.~English, A.J.~Sievers, B.~Ilic, D.A.~Czaplewski and H.G.~Craighead, Chaos {\bf 13}, 702 (2003).

\bibitem{SchwarzPRL1999} U.T.~Schwarz, L.Q.~English and A.J.~Sievers, Phys. Rev. Lett. {\bf 83}, 223 (1999).

\bibitem{EnglishPRE2008} L.Q.~English, R.~Basu Thakur and R.~Stearrett, Phys. Rev. E {\bf 77}, 066601 (2008).

\bibitem{swanson} B. I. Swanson, J. A. Brozik, S. P. Love, G. F. Strouse, A. P.
Shreve, A. R. Bishop, W.-Z. Wang, and M. I. Salkola, Phys.
Rev. Lett. {\bf 82}, 3288 (1999).

\bibitem{cuevas} J. Cuevas, L. Q. English, P. G. Kevrekidis, and M. Anderson
Phys. Rev. Lett. {\bf 102}, 224101 (2009)

\bibitem{BoechlerPRL2010} N.~Boechler, G.~Theocharis, S.~Job, P.G.~Kevrekidis, M.A.~Porter and C.~Daraio, Phys. Rev. Lett. {\bf 104}, 244302 (2010).

\bibitem{jinkyu} C. Chong, F. Li, J. Yang, M. O. Williams, I. G. Kevrekidis, P. G. Kevrekidis, and C. Daraio
Phys. Rev. E {\bf 89}, 032924 (2014)

\bibitem{fpuref} E. Fermi, J. Pasta, and S. Ulam, Tech. Rep. Los Alamos Nat.
Lab. LA1940 (1955)

\bibitem{Chaos} D. K. Campbell, P. Rosenau, and G. M.
Zaslavsky, Chaos {\bf 15}, 015101 (2005)

\bibitem{VK75} N.G. Vakhitov and A.A. Kolokolov,
Radiophys. Quantum Electron. {\bf 16} 783 (1973).

\bibitem{MS98} R.S. MacKay and J.-A. Sepulchre,
Physica D {\bf 119}, 148 (1998).

\bibitem{MAF98} J.L. Mar\'{\i}n, S. Aubry, and L.M. Flor\'{\i}a,
Physica D {\bf 113}, 283 (1998).

\bibitem{cls} T. Cretegny, R. Livi and M. Spicci, Physica D {\bf 119}, 88 (1998).

\bibitem{Noble} G. James and P. Noble, Physica D {\bf 196}, 124 (2004).

\bibitem{MA94} R.S. MacKay and S. Aubry,
Nonlinearity {\bf 7}, 1623 (1994).

\bibitem{Bambusi2} D. Bambusi,
Comm. Math. Phys. {\bf 324}, 515 (2013).


\bibitem{Archilla} J.F.R. Archilla, J. Cuevas, B. S\'{a}nchez-Rey, and A. \'{A}lvarez,
Physica D {\bf 180}, 235 (2003).

\bibitem{KK09} V. Koukouloyannis and P.G. Kevrekidis,
Nonlinearity {\bf 22}, 2269 (2009).

\bibitem{PelSak} D.E. Pelinovsky and A. Sakovich,
Nonlinearity  {\bf 25}, 3423 (2012).

\bibitem{CKP} J. Cuevas--Maraver, P.G. Kevrekidis, and D.E. Pelinovsky,
DOI: 10.1111/sapm.12107. Stud. Appl. Math. (2015).

\bibitem{fkm} S. Flach, K. Kladko and R.S. MacKay,
Phys. Rev. Lett. {\bf 78}, 1207 (1997).

\bibitem{bernardo} B. S\'anchez--Rey, G. James, J. Cuevas and J.F.R. Archilla,
Phys. Rev. B {\bf 70}, 014301 (2004).

\bibitem{Supp}
See the Supplementary Material at \texttt{[url]}, which includes Refs. \cite{PPP}, \cite{James1}, \cite{James2}, \cite{QinXiao}, \cite{PelSchn}, \cite{phason} and \cite{AMM99}, for (i) the derivation of expansion (\ref{dispersion}), (ii) an analysis of small-amplitude Klein-Gordon breathers, (iii) an extension of the result to FPU lattices, (iv) an energy criterion in the anti-continuous limit, and (v) a brief description of the numerical methods used for calculating discrete breathers.

\bibitem{PeyrardBishop} M. Peyrard and A.R. Bishop,
Phys. Rev. Lett. {\bf 62}, 2755 (1989).

\bibitem{cretegny} S. Aubry and T. Cretegny, Physica D {\bf 119}, 34 (1998).

\bibitem{Kastner} G. James and M. Kastner, Nonlinearity {\bf 20}, 631 (2007).

\bibitem{PPP} D. Pelinovsky, T. Penati, and S. Paleari, arXiv:1509.06389.

\bibitem{James1} G. James, J. Nonlin. Sci. {\bf 13} (2003), 27--63.

\bibitem{James2} G. James, B. S\'{a}nchez-Rey, and J. Cuevas, Rev. Math. Phys. {\bf 21} (2009), 1--59.

\bibitem{QinXiao} W.-X. Qin and X. Xiao, Nonlinearity {\bf 20} (2007), 2305--2317.

\bibitem{PelSchn} D.E. Pelinovsky and G. Schneider, arXiv:1603.05463.

\bibitem{phason} J. Cuevas, J.F.R. Archilla, and F.R. Romero, J. Phys. A: Math. Theor. {\bf 44}, 035102 (2011).

\bibitem{AMM99} J.F.R. Archilla, R.S. MacKay, and J.L. Mar\'{\i}n, Physica D {\bf 134}, 406 (1999).


\end{thebibliography}
\end{document}